\documentclass[twocolumn,prd,showpacs]{revtex4}

\begin{document}

\title{On the energy of homogeneous cosmologies}

\author{James M. Nester$^{1,2,3}$}
\email{nester@phy.ncu.edu.tw}
\author{Lau Loi So$^{1,4}$}
\email{s0242010@cc.ncu.edu.tw}
\author{T. Vargas$^3$}
\email{teovar@phy.ncu.edu.tw}
 \affiliation{$^1$Department of Physics,
 National Central University, Chungli 320, Taiwan}
 \affiliation{$^2$Graduate Institute of Astronomy,
 National Central University, Chungli 320, Taiwan}
  \affiliation{$^3$Center for Mathematics and Theoretical Physics,
National Central University, Chungli 320, Taiwan}
\affiliation{$^4$Current address: Department of Physics, Tamkang
University, Tamsui 251, Taiwan}

\date{\today}













\begin{abstract}
An energy for the homogeneous cosmological models is presented.
More specifically, using an appropriate natural prescription, we
find the energy within any region with any gravitational source for
a large class of gravity theories---namely those with a tetrad
description---for all 9 Bianchi types.   Our energy is given by the
value of the Hamiltonian with homogeneous boundary conditions; this
value vanishes for all regions in all Bianchi class A models, and it
does not vanish for any class B model.  This is so not only for
Einstein's general relativity but, moreover, for the whole
3-parameter class of tetrad-teleparallel theories. For the
physically favored one parameter subclass, which includes the
teleparallel equivalent of Einstein's theory as an important special
case, the energy for all class B models is, contrary to expectation,
{\em negative}.
\end{abstract}

\maketitle


\section{Introduction}
Gravity is the only universal force; it is long range and
dominates the cosmos. Energy has been one of the most useful
physical concepts, no less so in gravitating systems---one need
only recall its utility in the Newtonian Kepler problem. From the
modern relativistic perspective energy-momentum is the source of
gravity, yet---somewhat ironically---unlike all matter and other
interaction fields in the absence of gravity, identifying a good
description of the energy-momentum of gravitating systems has been
the oldest and most controversial outstanding puzzle. Physically
the fundamental difficulty can be understood as a consequence of
Einstein's equivalence principle---from which it follows that
gravity cannot be detected at a point. Hence the energy-momentum
of gravity cannot have a proper local density; it is fundamentally
non-local and thus inherently non-tensorial and therefore {\it
non-covariant}. (For a good discussion of this point see
~\cite{mtw}.)

After Einstein proposed his gravitational energy-momentum density
(see e.g.~\cite{tra}), various alternate prescriptions were
introduced by other researchers (notably Papapetrou~\cite{pap},
Bergman and Thompson~\cite{beto}, M{\o}ller~\cite{Mol58},
Goldberg~\cite{Gol58}, Landau and Lifshitz~\cite{lan}, and
Weinberg~\cite{wein}). These investigations lead to a variety of
expressions with no compelling criteria for favoring any particular
one. Moreover these traditional energy-momentum {\it pseudotensors}
are, as noted, necessarily not covariant objects, they inherently
depend on the coordinates, so they cannot provide a truly physical
local gravitational energy-momentum density. Caught between the
equivalence principle and the covariance principle, the pseudotensor
approach has been largely questioned, although never completely
abandoned.

On the other hand, because the gravitational interaction is local,
some kind of local---or at least nearly local---description of the
associated energy-momentum was still sought. The modern concept,
introduced by Penrose~\cite{Penrose} to resolve this dilemma,  is
that properly energy-momentum is {\it quasi-local}, being associated
with a closed surface bounding a region (for a nice review of the
topic see~\cite{Sza04}). However it soon became clear that there is
no unique quasi-local energy expression. Although there are some
especially famous ones (e.g., Brown-York~\cite{bryo} and
Liu-Yau~\cite{LY03}) many definitions of quasi-local mass-energy
have been proposed; they generally give distinct results. For
example Bergqvist~\cite{berg} studied several different quasi-local
mass definitions
 for the Kerr and Reissner-Nordstr\"{o}m spacetimes and came to the
conclusion that not even two of the examined definitions gave the
same result.

Our view is that from the Hamiltonian perspective one can make
sense of this situation---understanding not only why all these
otherwise perplexing choices exist but also what is their real
physical significance. Simply put the energy of a gravitating
system within a region---regarded as the value of the Hamiltonian
for this system---naturally depends not only on the interior of
the region but also on the {\em boundary conditions} imposed at
the interface with the exterior (this is reasonable, after all the
particular solution to the field equations depends on the boundary
conditions). It has been found that the Hamiltonian necessarily
includes an integral over the 2-surface bounding the considered
region.  This Hamiltonian boundary term plays two key roles: (i)
it controls the value of the Hamiltonian, and (ii) its specific
form (via the requirement that the boundary term in the variation
in the Hamiltonian vanish) is directly related to the selected
type and value of the boundary conditions. Many of the quasi-local
proposals (namely all those which admit a Hamiltonian
representation) can be understood in these terms: their
differences are simply associated with different boundary
conditions. Furthermore, using a covariant Hamiltonian formalism,
it has been shown~\cite{PRL99,GC00} that every energy-momentum
pseudotensor can be associated with a particular Hamiltonian
boundary term---which in turn determines the quasi-local
energy-momentum that is respectively linked with the implied
boundary conditions. In this sense, it has been said that the
Hamiltonian quasi-local energy-momentum approach rehabilitates the
pseudotensors and, moreover, dispels any doubts about the physical
meaning of these energy-momentum complexes---since with this
approach all their inherent ambiguities are given clear physical
and geometric meanings.

We want to emphasize that while there are many possible
Hamiltonian boundary term energy-momentum quasi-local
expressions---simply because there are many conceivable boundary
conditions---it nevertheless has been found~\cite{PRD05} that in
practice there usually is a particular choice {\em best suited} to
the task at hand. It has long been known that this is the case for
familiar physical systems where gravity is negligible. For example
in thermodynamics we have not a unique energy but rather the
internal energy, the enthalpy, and the Helmholtz and Gibbs free
energies---each a real physical energy adapted to a specific
interface between the physical system of interest and its
surroundings; similarly in classical electrostatics the physically
appropriate measure of energy obtained from the work-energy
relation for a finite system depends on whether one considers
fixed surface charge density or fixed potential---boundary
condition choices which are respectively associated with the
symmetric and the canonical energy-momentum tensor.

Here we apply this Hamiltonian boundary term quasi-local
energy-momentum approach to homogeneous cosmologies. Our
motivation is twofold. On the one hand the utility of having a
good measure of the energy of such gravitating systems---a measure
with sensible answers for the ideal exactly homogeneous case but
which can be applied to any perturbations thereof and even quite
generally---should be obvious. We are likewise motivated by the
consideration that homogeneous cosmology affords an excellent set
of models where one can test the suitability of our---or indeed
any other---proposed (quasi-)local energy-momentum ideas.

Accordingly it is important to here remark on our specific choice of
energy expression for these homogeneous cosmologies. We have noted
five different approaches which lead us to exactly the same formula
for energy-momentum for the homogeneous cosmologies in General
Relativity (GR). (i) One may take the tetrad form of GR: among the
many traditional energy-momentum approaches M{\o}ller's~\cite{Mol61}
tetrad-teleparallel formulation has been highly regarded. This form
of GR can be viewed
 from the gauge theory perspective, then a certain {\em
translational gauge current} (closely related to the energy-momentum
expression proposed by M{\o}ller) stands out (among other virtues it
is the only classical expression that has an associated {\em
positive energy} proof~\cite{Nes89}). (ii) Another perspective is
via the GR Hamiltonian: from the covariant Hamiltonian approach for
GR one particular Hamiltonian boundary term stands out as being the
favored expression for general applications~\cite{PRD05}. In many
situations, including the present application, it reduces to the
tetrad-teleparallel gauge current expression. (iii) Alternately one
can proceed from the spinor-parameterized Hamiltonian associated
with the Witten positive energy proof~\cite{Nes81,Nes84,NT94}. For
an isolated asymptotically flat gravitating system the spinor field
may be taken to satisfy Witten's equation, then the Hamiltonian has
a positive value; however for a {\em homogeneous} cosmology the
Witten equation does not have appropriate solutions, instead it is
appropriate to take the spinor field to be {\em homogeneous}, which
reduces the Hamiltonian to the aforementioned expression. (iv) A
fourth approach---which we like for its simplicity, generality and
straightforwardness---is to consider the general teleparallel theory
(which includes Einstein's GR as a special case). For such theories,
quite unlike the GR situation, investigators have advocated only two
specific expression for the energy-momentum density (this is one of
the virtues of treating this whole class of theories rather than the
one special case equivalent to GR), one of which---the
aforementioned tetrad-teleparallel gauge current---stands out as
most suitable for this and most other applications. (v) Instead one
could begin directly with homogeneous cosmology: homogeneous
cosmologies are naturally described in terms of a preferred
homogeneous tetrad; then an energy-momentum expression based on the
Hamiltonian for this preferred tetrad is clearly appropriate.

There have been several studies aimed at finding the total energy of
the expanding universe. An early investigation proposed that our
universe may have arisen as a quantum fluctuation of the vacuum.
That model predicted a universe which is homogeneous, isotropic and
closed, and consists equally of matter and anti-matter.
Albrow~\cite{Al73} and Tryon~\cite{try}  proposed that our universe
must have a zero net value for all conserved quantities and
presented some arguments, using a Newtonian order of magnitude
estimate, favoring the proposal that the net energy of our universe
may indeed be zero. The general argument for this requirement is
that energy can always be represented by an integral over a closed
2-surface bounding the region of interest, so if the universe has an
empty boundary then the energy should vanish (for a more detailed
discussion of the vanishing of energy for a closed cosmology
see~\cite{Tipler}). Thus for closed cosmological models the total
energy is necessarily zero. Years ago Misner~\cite{Mis63} pointed
out a technical problem with the attempts at explicit demonstrations
of this statement. He noted that the integrands that were being used
were reference frame dependent (holonomic) pseudotensors and their
associated superpotentials. None of those discussions had
specifically established exactly how these particular non-tensorial
objects behaved under changes of coordinates---but such changes were
actually necessary in the calculations since the whole universe
could not be covered with one coordinate patch. According to our
understanding it was Wallner~\cite{Wal82} who first gave for GR a
clear demonstration of this important vanishing energy for closed
universe requirement; his integrand (effectively the same
tetrad-teleparallel gauge current that we will use) was given in
terms of a globally defined frame field.

The subject of the total energy of expanding universe models was
re-opened by Cooperstock and Israelit~\cite{coop}, Rosen~\cite{ros},
Garecki~\cite{Gar95}, Jhori {\it et al.}~\cite{JKSE}, Feng and
Duan~\cite{FD96}, and others using various GR energy-momentum
definitions. In one of these investigations the Einstein
energy-momentum pseudotensor was used to represent the gravitational
energy~\cite{ros}, which led to the result that the total energy of
a closed Friedman-Lema\^{\i}tre-Robertson-Walker (FLRW) universe is
zero. In another, the symmetric pseudotensor of Landau-Lifshitz was
used~\cite{JKSE}. Some works calculated the total energy of certain
anisotropic Bianchi models using different pseudotensors, leading to
similar results~\cite{base,rad99,Xu00}. More recently Faraoni and
Cooperstock~\cite{facoo}, in support of Cooperstock's proposal that
gravitational energy vanishes in the vacuum, argued (with the aid of
a non-minimally coupled scalar source) that the open, or critically
open FLRW universes---as well as Bianchi models evolving into de
Sitter spacetimes---also have zero total energy. Finally,
calculations for the closed FLRW~\cite{var} and some anisotropic
Bianchi models~\cite{sovar} using the Einstein, Laudau-Lifshitz and
other complexes in teleparallel gravity also led to vanishing
energy.

In the present work, we examine the energy-momentum (for both
local and global regions) for a large class of cosmological
models---more specifically we consider all  9 Bianchi types of
homogeneous cosmological models using the Hamiltonian boundary
term quasi-local approach in the context of not only GR but more
generally the tetrad-teleparallel theory of gravity~\cite{HS79}.
This theory is a generalization of GR (it has been dubbed NGR, for
new general relativity); a certain special case which is
equivalent to Einstein's theory~\cite{Mol61,AGP00} was first
proposed by M{\o}ller to solve the energy localization problem.
This special case has been referred to as GR${}_{||}$, the
teleparallel equivalent to Einstein's theory (a.k.a.~TEGR);  it
has attracted the attention of several investigators, see
e.g.,~\cite{Nes89,KT91,Mal,AGP00,BV01,OR06}. Our motive for this
general approach is not only to improve our understanding of these
cosmological models, but also to better understand the considered
gravity theories and especially to better understand the meaning
and application of these quasi-local energy-momentum ideas.

Most of the previous calculations of the energy of the universe
used cartesian or spherical coordinate systems; here in contrast
we use only the symmetry of spacetime given by the Bianchi group.
In this major extension of previous results~\cite{var,sovar}, by
using the Hamiltonian formalism with {\it homogeneous boundary
conditions} we find that the energy---{\it for all regions} (any
shape, large or small)---{\it vanishes} for {\it all} Bianchi
class A models and {\it does not} vanish for any class B model.
This result does not depend on the specific type of matter content
of the universe (e.g., baryons, radiation, cold dark matter, dark
energy, cosmological constant); it depends only on the symmetry of
the spacetime given by the structure constants of the Bianchi
group. It is noteworthy that we find a simple physical difference
related to the division of the Bianchi types into two classes.
Moreover, for both GR and the physically viable teleparallel
theories, our energy for Bianchi class B turns out to be, contrary
to our expectations and presumptions, {\em negative}.

The outline of the rest of this work is as follows: in the next
section we discuss the Bianchi type universes.  In Section 3 we
briefly discuss the essential elements concerning the conserved
energy-momentum expressions for Einstein's general relativity (GR),
the general tetrad-teleparallel theory and the teleparallel
equivalent of GR. In Section 4 we give the respective
Hamiltonian-boundary-term quasi-local expressions. The energy
calculation is presented in Section 5.  This is followed by our
concluding discussion of the remarkable results.

\section{Bianchi type universes}\label{Bianchi}

Suppose that the four-dimensional spacetime manifold can be
foliated by homogeneous space-like hypersurfaces $\Sigma_t$ of
constant time $t$. Homogeneity means that each spatial
hypersurface has a transitive group of isometries. The study of
this 3-parameter isometry group, via the classification of
3-dimensional Lie algebras, led to the Bianchi classification of
spatially homogeneous universes~\cite{ellis}.

These models are characterized by homogeneous---but generally
anisotropic---spatial hypersurfaces parameterized by time. In a
synchronous coordinate systems, in which the time axis is always
normal to the hypersurfaces of homogeneity $\Sigma_t$, we take the
spacetime orthonormal (co)frame for these cosmological models to
have the form
\begin{equation}\vartheta^0={\rm d}t,\quad
\vartheta^a(t)=h^a{}_k(t)\sigma^k,\quad a=1,2,3\label{bcoframe}
\end{equation} where the three basis one-forms $\sigma^k=\sigma^k(x)$, $k=1,2,3$, depend on spatial
position in such a way that
\begin{equation}
{\rm
d}\sigma^k={\frac12}C^k{}_{lm}\sigma^l\wedge\sigma^m,\label{homo}
\end{equation}
with $C^k{}_{lm}$ being certain constants. The associated spacetime
metric then has the form
\begin{equation}
{\rm d}s^2=-{\rm d}t^2+g_{lm}(t)\sigma^l(x)\sigma^m(x),
\end{equation}
where $g_{lm}:=\delta_{ab}h^a{}_lh^b{}_m$ is a spatial 3-metric
which depends only upon time; for our analysis it need not be
diagonal.

There are 9 Bianchi types distinguished by the particular form of
the structure constants $C^k{}_{lm}$~\cite{ellis,kra}. They fall
into two special classes: class A (types I, II, VI${}_0$,
VII${}_0$, VIII, IX) have $A_k:=C^m{}_{km}\equiv0$ and class B
(types III, IV, V, VI${}_h$, VII${}_h$) are characterized by
$A_k\ne0$.  For our purposes here we hardly need any more details
regarding these types. We note that the respective scalar
curvatures are vanishing for Type I, positive for Type IX, and
negative for all the other types. Also, although the general idea
of these models is homogeneous but non-isotropic, certain special
cases can be isotropic; specifically, isotropic Bianchi I, V, IX
are, respectively, isometric to the usual
Friedmann-Lema\^{\i}tre-Robertson-Walker (FLRW) models:
$k=0,-1,+1$.

\section{GR and the tetrad-teleparallel theory}\label{theory}

Here we note some features, associated with conserved expressions
for energy-momentum, for Einstein's general relativity (GR) and
certain alternative tetrad-teleparallel theories, especially the one
which is equivalent to GR.

For all cases the analysis can be most conveniently expressed in
terms of differential forms and orthonormal (co)frames:
\begin{equation}
\vartheta^\alpha=e^\alpha{}_i\, {\rm d}x^i, \quad i=0,1,2,3.
\end{equation}
In such frames the metric coefficients are constant:
$g_{\mu\nu}=\mathrm{diag}(-1,+1,+1,+1)$. We will consider on the
same space more than one connection (equivalently, more than one
 covariant derivative or parallel transport rule). However, in all cases the connection is
assumed to be metric compatible:
\begin{equation}
0={\rm d}g_{\mu\nu}-\Gamma^\lambda{}_\mu
g_{\lambda\nu}-\Gamma^\lambda{}_\nu
g_{\mu\lambda}=-\Gamma_{\mu\nu}-\Gamma_{\nu\mu}.
\end{equation}
 Hence the connection one-form
coefficients are always antisymmetric:
$\Gamma_{\alpha\beta}=\Gamma_{[\alpha\beta]}$. Any particular
connection is characterized by its {\em curvature} (2-form):
\begin{equation}
R^\alpha{}_\beta={\rm
d}\Gamma^\alpha{}_\beta+\Gamma^\alpha{}_\gamma\wedge\Gamma^\gamma{}_\beta,
\end{equation}
and its {\em torsion} (2-form):
\begin{equation}
T^\alpha={\rm
d}\vartheta^\alpha+\Gamma^\alpha{}_\beta\wedge\vartheta^\beta.
\end{equation}

\subsection{GR}

Einstein's general relativity is based on a Riemannian geometry,
using the Levi-Civita connection---which has vanishing torsion. The
Einstein field equations can be expressed in the form
\begin{equation}
R^{\alpha}{}_{\beta}\wedge\eta_\alpha{}^\beta{}_\mu=-2\kappa {\cal
T}_\mu\,, \label{eineq}
\end{equation}
where
$\eta^{\alpha\beta\cdots}:=*(\vartheta^\alpha\wedge\vartheta^\beta\cdots)$
is the dual form basis. The expression on the left is the Einstein
3-form $-2G^\nu{}_\mu\eta_\nu$, and the quantity on the right is the
source energy-momentum 3-form $T^\nu{}_\mu\eta_\nu$, with
$\kappa=8\pi G/c^4$.  Here we take units such that $c=1$.

 Let us first note that the above Einstein
equation can naturally be rearranged into a certain special form
(for similar arguments see~\cite{Sza92,Nes04}):
\begin{equation}
{\rm d}(\Gamma^\alpha{}_\beta\wedge\eta_\alpha{}^\beta{}_\mu)=
-\Gamma^\alpha{}_\beta\wedge {\rm d}\eta_\alpha{}^\beta{}_\mu
-\Gamma^\alpha{}_\gamma\wedge\Gamma^\gamma{}_\beta\wedge\eta_\alpha{}^\beta{}_\mu-2\kappa{\cal
T}_\mu.\label{GRgaugecurrent}
\end{equation}
 The 2-form {\em
superpotential},
$\Gamma^\alpha{}_\beta\wedge\eta_\alpha{}^\beta{}_\mu$, is the key
object. Its differential is exact; consequently the 3-form on the
right hand side is closed; hence it is {\em automatically} a
conserved current---which includes the material energy-momentum
linearly. The remaining pieces on the right hand side (which depend
only on geometric quantities) can thus be interpreted as the
energy-momentum density of the gravitational field.  These forms are
just the quantities we wish to use to describe the energy of
GR---and this rearrangement of the field equation is the easiest way
to get them that we know of.

Nevertheless, here we want to also present some additional material
to (i) reinforce our thesis that these indeed are the most suitable
energy-momentum expressions for homogeneous GR cosmologies and also
to (ii) include a large class of alternate theories.

Hilbert noted that the above Einstein equation (\ref{eineq}) can be
obtained from the scalar curvature Lagrangian density by regarding
it as a function of the metric. Later certain variations on the
theme were considered. For the vacuum---or even non-derivative
coupled sources like scalar fields or the Maxwell and Yang-Mills
gauge fields---one can simply vary the coframe and the metric
compatible connection one-form independently. Then the connection
variation yields an equation which requires the connection to have
vanishing torsion, i.e., to be the Levi-Civita connection of GR. Let
us next note a result~\cite{gensource}
 we can get most simply by proceeding from that
Lagrangian formulation.

 The scalar curvature Lagrangian 4-form, ${\cal L}_{\rm R}=R\eta=R^{\alpha\beta}\wedge\eta_{\alpha\beta}$, can be rearranged as indicated:
\begin{eqnarray}
R^\alpha{}_\beta\wedge\eta_\alpha{}^\beta
\equiv\left({\rm d}\Gamma^\alpha{}_\beta+\Gamma^\alpha{}_\gamma\wedge\Gamma^\gamma{}_\beta\right)\wedge\eta_\alpha{}^\beta\equiv \qquad&&\nonumber\\
{\rm
d}\left(\Gamma^\alpha{}_\beta\wedge\eta_\alpha{}^\beta\right)+\Gamma^\alpha{}_\beta\wedge
{\rm d}\eta_\alpha{}^\beta
+\Gamma^\alpha{}_\gamma\wedge\Gamma^\gamma{}_\beta\wedge\eta_\alpha{}^\beta.&&
\end{eqnarray}
Now by dropping the total differential we obtain
\begin{equation}
2\kappa{\cal L}_{\mathrm{ tet1}}=-{\rm
d}\vartheta^\mu\wedge\Gamma^\alpha{}_\beta\wedge
\eta_\alpha{}^\beta{}_\mu
+\Gamma^\alpha{}_\gamma\wedge\Gamma^\gamma{}_\beta\wedge\eta_\alpha{}^\beta,\label{Lprime}
\end{equation}
a modified Lagrangian density which gives the same field equations.
It is noteworthy that the canonical momentum conjugate to the
co-frame,
\begin{equation}
\tau_\mu:=\frac{\partial{\cal L}_{\mathrm{tet1}}}{\partial {\rm
d}\vartheta^\mu}=-\Gamma^{\alpha\beta}\wedge \eta_{\alpha\beta\mu}
=-\Gamma^{\alpha\beta}{}_\gamma\delta^{\tau\rho\gamma}_{\alpha\beta\mu}\frac12\eta_{\tau\rho}\,,\label{momflux}
\end{equation}
is just the already encountered superpotential 2-form.

Recall that the Levi-Civita connection is linear in the differential
of the frame:
\begin{equation}
\Gamma_{\alpha\beta}= \frac12\left[({\rm
d}\vartheta_\alpha)_{\beta\gamma}+({\rm
d}\vartheta_\beta)_{\gamma\alpha} +({\rm
d}\vartheta_\gamma)_{\beta\alpha}\right]\vartheta^\gamma\,.\label{levicivita}
\end{equation}
This relation can be used to eliminate $\Gamma^\alpha{}_\beta$ from
(\ref{Lprime}) to obtain a certain specific action: ${\cal L}_{\rm
GRtet}={\cal L}_{\rm GRtet}(\vartheta,{\rm d}\vartheta)$, which is
quadratic in ${\rm d}\vartheta$. In addition to this tetrad version
of Einstein's theory there are other more general tetrad theories of
gravity which follow from other Lagrangians quadratic in ${\rm
d}\vartheta$. Such tetrad theories are somewhat interesting in their
own right. Let us consider them in the next subsection.

\subsection{tetrad-teleparallel theory}
In this work we are concerned especially with the general
homogeneous Bianchi cosmologies.  Such models have a natural
preferred global homogeous orthonormal frame (a.k.a.~tetrad,
vierbein). If a geometry has a preferred tetrad one can naturally
introduce a new parallel transport rule (i.e., a new  connection)
such that this frame is, by definition, {\em parallel}.  A geometry
with a preferred global parallelism is referred to as {\em
teleparallel\/} (a.k.a.~{\em absolute parallel\/}).  Hence,
especially for the homogeneous cosmologies, an appropriate
theoretical geometrical framework is the tetrad/teleparallel
formulation.

Conversely, for any teleparallel geometry {\em by definition} the
curvature vanishes and the parallel transport is path independent.
It follows that a global preferred frame may be constructed by
starting with any orthonormal frame at a single
 point and parallel transporting it along any path to all the other
points.  The resultant global tetrad field is unique (up to an
overall global constant Lorentz transformation).  In this
constructed frame the connection coefficients vanish. Hence in a
teleparallel geometry there is (up to a an overall global constant
rotation) a {\em preferred frame} in which the connection also
vanishes. The basic variable can then just be taken to be this
preferred tetrad, which is most conveniently represented as the
(co)frame $\vartheta^\alpha=e^\alpha{}_i\, {\rm d}x^i$.
 Since the connection coefficients vanish in this frame the
associated teleparallel torsion 2-form is then simply given by the
frame differential:
\begin{equation}
T^\alpha=\frac12
T^\mu{}_{\alpha\beta}\vartheta^\alpha\wedge\vartheta^\beta={\rm
d}\vartheta^\alpha\,. \label{torsion}\end{equation} Thus we can
forgo further mention of the teleparallel connection here and simply
regard a teleparallel theory as a theory for a preferred tetrad.

Tetrad theory field equations can be obtained from a Lagrangian
4-form ${\cal L}_{\rm{tot}}={\cal
L}_{\rm{tet}}(\vartheta^\alpha,{\rm d}\vartheta^\alpha)+{\cal
L}_{\rm{mat}}(\vartheta)$. The variation with respect to the frame
gives
 \begin{equation}
 \delta{\cal L}_{\rm{tot}}= \delta{\rm d}\vartheta^\mu\wedge
 \frac{\partial{\cal L}_{\rm{tet}}}{\partial
 {\rm d}\vartheta^\mu}+\delta\vartheta^\mu\wedge\left(\frac{\partial{\cal
 L}_{\rm{tet}}}{\partial\vartheta^\mu}+\frac{\partial{\cal
 L}_{\rm{mat}}}{\partial\vartheta^\mu}\right).
\end{equation}
This expression has the form
\begin{eqnarray}
\delta{\cal L}_{\rm{tot}} =
 \delta {\rm d}\vartheta^\mu\wedge\tau_\mu
 +\delta\vartheta^\mu\wedge (-t_\mu-{\cal T}_\mu)\equiv && \nonumber\\
{\rm
d}\left(\delta\vartheta^\mu\wedge\tau_\mu\right)+\delta\vartheta^\mu\wedge({\rm
d}\tau_\mu-t_\mu-{\cal T}_\mu),&& \label{tetfe}\end{eqnarray}
which identifies  $\tau_\mu$ as the canonical conjugate field
momentum and $t_\mu$ and ${\cal T}_\mu$ as, respectively, the
3-forms of gravitational energy-momentum and the material source
energy-momentum density. Using (\ref{tetfe}) and Hamilton's
principle gives the tetrad field equation in a form reminiscent of
that used by Einstein in his search for his gravity field
equations~\cite{JR06}:
\begin{equation}
{\rm d}\tau_\mu=t_\mu+{\cal T}_\mu.\label{gaugecurrent3}
\end{equation}
The right hand side is {\em naturally} a conserved current 3-form:
\begin{equation}
{\rm d}(t_\mu+{\cal T}_\mu)=0,
\end{equation}
(i.e., because the lhs is exact the rhs must be closed). The
associated conserved total energy-momentum within a volume $V$ is
\begin{equation}
P_\mu(V)=\int_V t_\mu+{\cal T}_\mu=\int_V {\rm
d}\tau_\mu=\oint_{\partial V} \tau_\mu.\label{qlem}
\end{equation}
The tetrad-teleparallel formulation is natural from the {\em gauge
theory} point of view.  From that perspective the above
expressions are those of the {\em translational gauge current}
(the conserved current associated with spacetime translations
according to Noether's first theorem, see,
e.g.,~\cite{Nes89,KT91,Mal,AGP00,BV01,Nes04}).

For the tetrad-teleparallel theories the canonically conjugate
momentum field,
\begin{equation}\tau_\mu:=\frac{\partial{\cal
L}} {\partial
T^\mu}={\frac12}\tau_\mu{}^{\alpha\beta}\eta_{\alpha\beta},
\end{equation}
 is generally
taken (in order to have quasi-linear second order field equations)
to be a linear combination,
\begin{equation}
\tau =\kappa^{-1}(a_1T_{\rm ten}+a_2T_{\rm vec}+a_3T_{\rm axi}),
\end{equation}
 of the tensor, vector, and axivector
irreducible parts of the teleparallel torsion~\cite{Hehl}:
\begin{eqnarray}
T_{\rm
vec}{}^\alpha{}_{\mu\nu}&:=&\frac23\delta^\alpha{}_{[\nu}T^\lambda{}_{\mu]\lambda},\quad
T_{\rm
axi}^{\alpha\mu\nu}:=\frac1{3!}\delta^{\alpha\mu\nu}_{\lambda\sigma\kappa}T^{\lambda\sigma\kappa},\nonumber\\
 T_{\rm ten}&:=&T-T_{\rm vec}-T_{\rm axi}.
\end{eqnarray}

 There is
thus a 3-parameter class of such theories~\cite{HS79}. The generic
theory determines a {\em preferred frame}.  It turns out, however,
that one special parameter choice ($4a_3=a_2=-2a_1$) is
distinguished: it actually has {\em local Lorentz gauge freedom}.
This model, with  $a_1=-1$, has been encountered
hereinbefore~(\ref{momflux}); it is known as GRtet or GR${}_{||}$,
the teleparallel equivalent of Einstein's GR (a.k.a.~TEGR), and
was first proposed by M{\o}ller~\cite{Mol61} to solve the GR
energy localization problem.

\section{Quasi-local boundary expression}
\label{Hamiltonian}

As mentioned in the introduction, for Einstein's GR many
energy-momentum expressions---both quasi-local and reference frame
dependent pseudotensors---have been proposed. It should be
emphasized that, despite much effort and many nice results, there
is {\it no consensus} as to which, if any, is best. The
Hamiltonian approach certainly helps. From that perspective the
energy-momentum is determined by the boundary term in the
Hamiltonian~\cite{CQG99,PRL99}. Although (at least formally) the
formalism allows for an infinite number of Hamiltonian boundary
expressions (including all the superpotentials that generate the
pseudotensors), the ambiguities have been tamed: each expression
has a geometrically and physically clear significance associated
with the boundary conditions  determined from the variation of the
Hamiltonian~\cite{Nes04}. Nevertheless there are (at least
formally) an infinite number of possible boundary conditions.

\subsection{Tetrad-teleparallel energy-momentum}
On the other hand, the situation for the tetrad/teleparallel theory
is in sharp contrast.
Investigators~\cite{Mol61,HS79,Wal82,Nes89,KT91,Mal,Mal+,AGP00,BV01,Mal+02}
were led to only two \cite{exceptions} closely related quasi-local
boundary term expressions for the energy-momentum within a volume
$V$:
\begin{equation}P_i(V):=\oint_{\partial V} e^\mu{}_i\tau_\mu, \qquad
P_\mu(V):=\oint_{\partial V} \tau_\mu,
\end{equation}
they are, respectively, the M{\o}ller 1961~\cite{Mol61} expression
and the {\em translational gauge current} derived above. M{\o}ller
had pointed out that his superpotential (which appears here as a
2-form integrand) is {\em tensorial} (i.e., it transforms
homogeneously under a change of coordinates); however its
differential,
\begin{equation}
{\rm d}(e^\mu{}_i \tau_\mu)={\rm
d}e^\mu{}_i\wedge\tau_\mu+e^\mu{}_i\, {\rm d}\tau_\mu,
\end{equation}
 the M{\o}ller tetrad-teleparallel
energy-momentum 3-form, {\em is not a tensor} with respect to
coordinate transformations (as M{\o}ller himself noted)---because of
the factor ${\rm d}e^\mu{}_i$. In contrast, it should be emphasized
that both the translation gauge current superpotential 2-form
$\tau_\mu$ and its differential,
 the gauge current 3-form (\ref{gaugecurrent3}), {\em are true tensors} under
changes of {\em coordinates}. Generically, the tetrad-teleparallel
theory has a natural preferred frame (no local frame gauge freedom),
then the translational gauge current energy-momentum expressions
have no ambiguity at all.

However for the one special case of greatest interest, GRtet, the
theory does have local Lorentz gauge freedom. In that case the gauge
current expressions do depend on the choice of orthonormal frame,
and thus still contain some observer dependent information mixed in
with the physical information in the energy-momentum expression.
Nevertheless we can regard the gauge current expressions as
preferable to any of the pseudotensors or M{o}ller's 1961
expression, since dependence on an orthonormal frame is more
physical than dependance on an arbitrary choice of coordinates.

Concerning the ambiguity of the choice of frame for GRtet, it is
important to note that the {\em quasi-local values} depend only on
the choice of frame on the boundary, and not on the choice within
the interior of the region.  Moreover in the case of interest here
(homogeneous cosmologies) there {\em is\/} a preferred frame---and
thus there is {\em no ambiguity} at all.

In summary, in the tetrad-teleparallel formulation unlike GR there
is no big ambiguity in the choice of expression. Thus our
consideration of this general tetrad/teleparallel class of theories
yields two benefits: (i) it allows us to get a result of great
generality, applying to this whole class of theories, (ii) moreover,
when specialized to TEGR, it determines a {\it unique} preferred
energy-momentum expression for GR.  This expression, the
tetrad-teleparallel gauge current has been regarded as one of the
best, perhaps the best, description of the gravitational
energy-momentum for GR.

\subsection{The covariant Hamiltonian approach}

On the other hand, let us also briefly discuss the alternative of
taking the usual Riemannian geometry approach to GR.  A covariant
Hamiltonian formulation has been developed for general geometric
gravity theories.  The Hamiltonian which generates the evolution of
a spatial region $V$ along the vector $N$ is given by an integral of
the form
\begin{equation}
H(N,V)=\int_V N^\mu{\cal H}_\mu +\oint_{\partial V} {\cal B}(N).
\end{equation}
It turns out that ${\cal H}_\mu$ is proportional to field
equations and thus has vanishing value (e.g., for GR (\ref{eineq})
we have ${\cal
H}_\mu=-(2\kappa)^{-1}R^\alpha{}_\beta\wedge\eta_\alpha{}^\beta{}_\mu-{\cal
T}_\mu$). The Hamiltonian boundary term controls the Hamiltonian
value and the boundary conditions. It has considerable freedom.
The analysis led to certain specific quasi-local Hamiltonian
boundary term 2-form expressions related to various types of
boundary condition choices (e.g., Dirichlet,
Neumann)~\cite{CQG99,GC00,Nes04} for quite general gravity
theories. When specialized to Riemannian GR, it was noted that one
of these expressions is singled out, in that it corresponds to
boundary conditions imposed on a complete 4-covariant object (the
tetrad) and gives the desired Bondi energy flux~\cite{PRD05}:
\begin{equation}
2\kappa{\cal B}(N)=\Delta\Gamma^\alpha{}_\beta\wedge
i_N\eta_\alpha{}^\beta+{\bar D}_\beta N^\alpha \Delta
\eta_\alpha{}^\beta.
\end{equation}
Here $N$ describes a spacetime vector field which selects the
components of energy and momentum, $\Delta\Gamma:=\Gamma-\bar\Gamma$
and $\Delta\eta:=\eta-\bar\eta$, where $\bar\Gamma$ and $\bar\eta$
are reference (or ground state) values.  (We note that the same
energy-momentum expression in its holonomic form was found by Katz,
Bi\v{c}{\'a}k, and Lynden-Bell. They have extolled its virtues in
several works, see in particular~\cite{KBL97} and the references
therein.)

Appropriate choices for the Homogeneous cosmologies are to take the
frame to be the preferred Bianchi frame, the reference connection to
have vanishing components in this frame, and for the spacetime
displacement vector field $N$ to be homogeneous, i.e., to have
constant components in this frame. Consequently the second term
vanishes and the boundary expression reduces to
\begin{equation}
2\kappa{\cal B}(N)=\Gamma^{\alpha\beta}\!\wedge
i_N\eta_{\alpha\beta}=N^\mu
\Gamma^{\alpha\beta}{}_\gamma\delta_{\alpha\beta\mu}^{\gamma\tau\rho}\frac12\eta_{\tau\rho}.\!\!\label{GRB}
\end{equation}
This is, as promised, the same superpotential form encountered
earlier (\ref{GRgaugecurrent},\ref{momflux}), an expression whose
utility has been recognized at least since~\cite{Nes81}. This
succinct argument
 shows that the preferred Riemannian GR covariant Hamiltonian quasi-local boundary
term coincides in this situation with the GR${}_{\|}$ gauge
current. The same expression also follows naturally from the
Hamiltonian formulation of the tetrad-teleparallel
theory~\cite{Nes89} and, moreover, from the Hamiltonian boundary
term associated with the Witten positive energy
proof~\cite{Nes81,Nes84,NT94} with homogeneous spinor field.

\section{The Bianchi energy calculation} \label{energy}

As we have argued, one can regard GR as a special case of the
tetrad-teleparallel theory, so we consider in detail the latter
formulation. The energy-momentum integral over the boundary of a
region at a fixed time $t$ is
\begin{equation}P_\mu(V):=\oint_{\partial V} \tau_\mu= \oint_{\partial
V}{\frac12}\tau_\mu{}^{\alpha\beta}\eta_{\alpha\beta}.
\end{equation}
In this Bianchi cosmology case the components of
$\tau_\mu{}^{\alpha\beta}$ and $T^\mu{}_{\alpha\beta}$ (in the
preferred teleparallel frame) are functions of time
alone---dependence on the spatial coordinates shows up only in the
teleparallel coframe $\vartheta^\mu$ via the $\sigma^m$.
Consequently, in detail
\begin{eqnarray}
P_\mu(V)&=&{\frac12}\tau_\mu{}^{\alpha\beta}(t)\oint_{\partial
V}\eta_{\alpha\beta}={\frac12}\tau_\mu{}^{\alpha\beta}(t)\int_{
V}{\rm
d}\eta_{\alpha\beta}\\
&=&{\frac12}\tau_\mu{}^{\alpha\beta}(t)\int_{
V}{\rm d}\vartheta^\gamma\wedge\eta_{\alpha\beta\gamma}
\nonumber\\
&=&{\frac12}\tau_\mu{}^{\alpha\beta}(t){\frac12}T^\gamma{}_{\lambda\delta}(t)
\delta^{\kappa\lambda\delta}_{\alpha\beta\gamma}\int_{
V}\eta_\kappa\nonumber\\&=& {\frac12}\tau_\mu{}^{\alpha\beta}(t)[
T^0{}_{\alpha\beta}+T^\gamma{}_{\gamma\alpha}\delta^0_\beta
-T^\gamma{}_{\gamma\beta}\delta^0_\alpha](t) V.\nonumber
\end{eqnarray}
Note that the energy-momentum is given by an integral over the
2-dimensional boundary of the region, yet it turns out to be simply
proportional to the size of the included 3-dimensional volume (not
its shape); it is also noteworthy that there is no dependence on the
location (this is just as it should be considering homogeneity).
Effectively, because of the {\em homogeneity}, we get a homogeneous
energy-momentum density, and a unique {\em localization} of
energy-momentum.

To properly appreciate our results here, one should note that most
previous calculations of energy for cosmological models considered
the isotropic FLRW models, or one or at most a few Bianchi types
(and they then usually confined their results to diagonal metrics).
They generally used {\it holonomic} energy-momentum expressions,
which correspond to rather different boundary conditions. Typically
they considered some ball of constant radius around the origin or
only the whole space. The expressions were not manifestly
homogenous; the results were shape dependent, not simply
proportional to volume.

Continuing with our calculation, we need to find the explicit
value of the teleparallel torsion (\ref{torsion}). From
(\ref{bcoframe}) we find
\begin{equation}
T^0=0,\quad T^a={\rm d} (h^a{}_k \sigma^k)=\dot h^a{}_k\, {\rm
d}t\wedge \sigma^k+h^a{}_k\, {\rm d}\sigma^k.
\end{equation}
using this along with (\ref{homo}) we obtain the torsion tensor
components $T^0{}_{\mu\nu}=0$ and
\begin{equation}
 T^a{}_{0 b}=\dot h^a{}_k h^k{}_b, \quad T^a{}_{bc}=h^a{}_k
C^k{}_{lm}h^l{}_b h^m{}_c.
\end{equation}
From the latter it follows that
\begin{equation}
 T^\gamma{}_{b\gamma}=T^c{}_{bc}=C^m{}_{km}h^k{}_b=A_k h^k{}_b\,.
\end{equation}
Consequently the energy-momentum is
\begin{eqnarray}
P_\mu(V)&=&\tau_\mu{}^{\alpha\beta}(t)
T^\gamma{}_{\beta\gamma}(t)\delta^0_\alpha V
\\
&=&\tau_\mu{}^{0b}(t) T^c{}_{bc}(t)V=\tau_\mu{}^{0b}(t)
A_kh^k{}_b(t)V ,\nonumber
\end{eqnarray}
which vanishes for all class A models.  (The physical interpretation
is that the negative gravitational binding energy density exactly
cancels the positive material energy density.)

For class B models we need
\begin{eqnarray}\kappa\tau_\mu{}^{0b}\!\!\! &=&\!\!\!  (a_1 T_{\rm ten} + a_2 T_{\rm vec} +a_3
T_{\rm axi})_\mu{}^{0b}\nonumber\\
&=&\!\!\!\left[a_1 T + (a_2-a_1) T_{\rm vec} + (a_3-a_1)T_{\rm
axi}\right]_\mu{}^{0b}\,.
\end{eqnarray}
 For energy we need just the $\mu=0$ component.  As $T^0{}_{\mu\nu}=0$ and $T_{\rm axi}$ is totally antisymmetric it reduces to
\begin{eqnarray}\kappa\tau_0{}^{0b}&=&
(a_2-a_1)T_{\rm
vec}{}_0{}^{0b}=(a_2-a_1){\frac13}T^c{}_c{}^b\nonumber\\
&=&(a_2-a_1){\frac13} A_k h^k{}_a \delta^{ab}.
\end{eqnarray}
Hence the energy within any volume $V$ at time $t$ is
\begin{equation}
P^0(V)=g^{00}\tau_0{}^{0b}(t) A_kh^k{}_b(t)V=\frac{a_1-a_2}{3\kappa}
A_k A_l g^{kl}(t)V.
\end{equation}

From various investigations it has been found that certain
parameter restrictions should be imposed. In particular for the
proper Newtonian limit we must require the so-called {\it viable}
condition~\cite{Hehl}, $2a_1+a_2=0$. Moreover normalization to the
Newtonian limit gives $a_1=-1$. This leads us to the one parameter
teleparallel theory, also known as NGR (New General
Relativity~\cite{HS79}). If these parameter conditions are
satisfied we have, for all viable cases (including the very
special case of GR${}_{\|}$, for which we also have $a_1+2a_3=0$),
\begin{equation}
E=P^0(V)=-\kappa^{-1}A_kA_lg^{kl}(t)V(t)<0,
\end{equation}
i.e., {\em negative energy} for all regions in class B models. With
similar computations these same energy results can be directly
verified for GR using the relations
(\ref{momflux},\ref{levicivita},\ref{qlem},\ref{GRB}).

Similarly one could find the explicit value of the ``linear
momentum'' $P_c(V)$ \cite{linearmom}. However, to get this one needs
to evaluate
\begin{equation}\tau_c{}^{0b} =\kappa^{-1}\left[a_1 T + (a_2-a_1) T_{\rm vec} +
(a_3-a_1)T_{\rm axi}\right]_c{}^{0b},
\end{equation}
which is linear (but not so simple) in $T^c{}_{0b}={\dot h}{}^c{}_j
h^j{}_b$.  The calculation is straightforward but not enlightening.

\section{Discussion and conclusions}

We have obtained some new insight regarding the energy of
homogeneous cosmologies, especially from considering the questions:
what energy should be associated with a region of the universe? Does
the total energy of a closed universe vanish? Is the energy of an
open universe positive? Does the energy of empty flat space vanish?

Specifically, using a natural prescription we found the value of the
(quasi-)local energy-momentum for the general tetrad-teleparallel
theory (which includes Einstein's GR as an important special case)
for all 9 Bianchi types---with general homogeneous gravitational
sources. From the Hamiltonian approach we found that, for comoving
observers with {\it homogeneous} boundary conditions and reference,
the energy vanishes for all regions in all Bianchi class A models,
and it does not vanish for any class B model. This is the case for
the whole 3-parameter class of tetrad-teleparallel theories.
According to our measure, the one parameter set of viable
teleparallel theories with a good Newtonian limit, which includes
the teleparallel equivalent of Einstein's theory, has {\em negative}
energy for all class B models.

We note that all the cosmologies in the Bianchi class A models can
be compactified, so our vanishing energy result is {\it
consistent} with the requirement of vanishing total energy for a
closed universe. The class B models cannot in general be
compactified, however there are some special exceptions.
Nevertheless these exceptions are not counter-examples: while in
certain cases the metric geometry can be compactified, this cannot
be done in such a way that the {\it frames} match up to give a
globally defined smooth frame (i.e., they are not {\em globally}
homogeneous). Indeed a proof of this~\cite{AS91} used a
calculation virtually identical to ours above without noting the
energy interpretation.

It is noteworthy that our energy depends only on the symmetry of the
spacetime given by the structure constants of the Bianchi group, and
that our result holds for {\it all} types of material sources
including {\it dark matter} and {\it dark energy} (either as a
cosmological constant or as some kind of unusual field like {\it
quintessence} appearing as a part of the energy-momentum tensor).

The results presented here can be specialized to the few Bianchi
models which can be isotropic.   For those cases, our {\it
homogenous} results have been compared with those found using a
similar approach for the more familiar
``isotropic-about-one-point'' FLRW
formulations~\cite{MPLA7,NCL07}. The isotropic Bianchi I is
isometric to the flat $k=0$ model, and both are found to have
vanishing energy.  The isotropic Bianchi IX is isometric to the
$k=+1$ model; the energy of the latter for spherically symmetric
regions first increases and then decreases to zero as the radius
reaches the antipode so that the volume encompasses the whole
close space (thereby satisfying the vanishing energy for any
closed space criteria), whereas the Bianchi IX has zero energy for
all regions. The open $k=-1$ FLRW model was found to have negative
energy, qualitatively but not quantitatively like its isometric
Bianchi counterparts V and VII$_h$.  It is remarkable that one
special case, with scale factor $a(t)=t$, is actually isometric to
empty Minkowski space with the expanding spatial slicing,
$t=\sqrt{T^2-R^2}$. The energy is {\em negative} for our measure
applied to the expanding space but, of course, the energy vanishes
for Minkowski with the constant $T$ slicing.  Thus the open and
closed models offer different ``localizations of energy'' for the
same physical situation, and provide  good examples of the effect
of different boundary conditions and time evolution vectors.

Although our energy values are obtained as integrals over the
boundary of a region, they are not truly {\em quasi-local}. Properly
a quasi-local quantity depends only on the physical data on the
boundary of a region.  Our values are obtained from integrals of the
Hamiltonian boundary term over the boundary, {\em but} the boundary
values, the reference values, and the displacement vector $N$ are
selected using the {\em global} homogeneity.  This gives us {\em
homogeneous localizations}.  For these models the homogeneity gives
a physically meaningful preferred {\em localization}.

 Suppose one were to take a region
of a homogeneous cosmology and regard it as an {\em isolated}
gravitating system with the exterior being empty, then far away the
isolated system would act like a Newtonian mass point and the
asymptotic gravitating field should be nearly static. Then there are
strong fundamental arguments
(see~\cite{ADM60,BD68,Ger78,BJ80,Hor84}) that the total effective
mass-energy---physically determined by the parameters of a large
Kepler orbit but mathematically defined as a certain 2-surface
integral at spatial infinity---should be positive (i.e., gravity is
attractive), and it should vanish iff the system is empty Minkowski
space. In particular an isolated gravitating system would allow
energy to be extracted---or if left alone it would spontaneously
radiate---until it reached the lowest available energy state.  By
scaling the lowest energy cannot have a finite negative value. A
lower bound of negative infinity allows one to extract an infinite
amount of energy, a source of perpetual motion, contrary to a
fundamental thermodynamics principle---such a system would never be
stable, it would continue to radiate forever. Thus the lowest state
should be non-negative. The ground state of geometric gravity is
Minkowski space with zero energy.

From this perspective it is reasonable to expect that the {\em
quasi-local energy} determined by a suitable boundary integral at
a finite radius, would also be non-negative and would vanish iff
the interior were flat Minkowski space; indeed these properties
have been regarded as desiderata for any good quasi-local energy
(see, e.g.,~\cite{Sza04,LY06}).

The measure of energy used here for homogeneous cosmologies is based
on a different concept.  Here we regard our cosmological region not
as an isolated system but rather as existing in an exterior which is
 just like the interior.  This is the essential significance of our
choice of homogeneous boundary conditions, reference and evolution.
One consequence is that our cosmological energies do not satisfy the
two important aforementioned quasi-local desiderata: indeed for the
expressions considered {\em positivity} need not hold, and {\em zero
energy} iff flat Minkowski space does not hold---in either
direction.

Regarding $E = 0$. For any {\it homogeneous} measure of energy in
Bianchi class A models it is certainly quite reasonable to have a
vanishing value for all regions, since a homogeneous energy density
energy must necessarily vanish at least for all compactifiable
regions---and these models can be compactified (this is most obvious
for Bianchi I with 3 torus topology identifications): the energy of
a closed universe must vanish.  This does not conform to the
standard {\em quasi-local} criterion of a unique quasi-local $E=0$
Minkowski ground state.

 The most remarkable feature of
our results is that many cosmological models were discovered to have
{\em negative} energy.
It had been expected that we would always find {\em non-negative}
energy. One consequence of this strong belief was that a minor
sign error was, quite unfortunately, overlooked in some
calculations; this lead to an {\em incorrect} claim of {\em
positive energy} for Bianchi B models being reported in some
recent conference proceedings~\cite{MPLA7,NCL07}.

It should be noted that a negative energy value for these open
Bianchi cosmologies is not just a peculiar feature due to our choice
of expression or model. Indeed the results of calculations (reported
in the aforementioned conference proceedings), done using
essentially the same principles articulated here, likewise led to
{\em negative energy} for the open FLRW (homogenous and isotropic)
model. Furthermore, according to our simple calculations, most
commonly used energy expressions will give the same energy signature
for both the open FLRW and Bianchi cosmologies. Yet, remarkably, we
have not found any conspicuous reports of this simple fact, although
there is some evidence which implicitly supports our energy
signature. In particular Bannerjee and Sen~\cite{base} evaluated the
Einstein pseudotensor for some Bianchi models.  Among other results,
they reported a non-vanishing energy for finite regions in some
class B models (specifically, types III, V, VI), but they did not
remark that the sign (easily evaluated) of their calculated energy
value for these models is actually negative. The Landau-Lifshitz
pseudotensor was used by Jhori {\it et al.}~\cite{JKSE} to calculate
the energy for the closed $k=+1$ FLRW model; one can see from their
Eq.~(10) that they would have found a negative value for the $k=-1$
model.

The stability argument for non-negative energy is quite
compelling---for isolated gravitating systems, which should be
settling down to a stable equilibrium state. The cosmological models
considered here, however, are quite different in kind. They are
inherently {\em dynamic}, very unlike stable isolated systems. We do
not see that there is any fundamental objection to assigning to them
a negative energy value.

Moreover, according to the following argument it is quite reasonable
that negative spatial curvature models may have negative energy.
Consider that a region of positive spatial curvature is like a
convex gravitational lens, tending to focus light rays, acting as if
it had a positive matter density which attracted the rays.  Whereas
a negative spatial curvature region acts like a concave lens,
causing light rays to be defocused, acting as if it had a negative
matter density which repelled light rays.  Our definition of energy
is actually the value of the Hamiltonian but this can be expected to
have some correlation with the effective active gravitational mass
of the region.

Identifying a good measure for the energy of a gravitating system
has remained an outstanding problem.  Here we considered this issue
for any region in homogeneous cosmologies. We presented for all
models the energy value associated with appropriate boundary
conditions as given by our favored covariant
Hamiltonian-boundary-term approach. We discovered, in contrast to
the expectations, zero energy for some non-flat cases and, most
surprisingly, negative energy for many (all Bianchi Class B).

Finally, let us again emphasize that our specific result hinges on
the particular chosen measure of energy-momentum.  We noted five
perspectives which led to the same energy-momentum expression for
GR:  certainly the chosen expression is a natural one for both
homogenous cosmologies and the teleparallel theories, and it
coincides with the GR$_{||}$ tetrad-teleparallel gauge current.
Yet our most fundamental argument for distinguishing this
energy-momentum expression is via the covariant Hamiltonian
approach---which includes {\em and goes beyond} the Noether
analysis~\cite{Nes04}. It should be noted that we apply the
Hamiltonian analysis to the general theory, and then impose the
homogeneous symmetry on the resulting expressions. It has long
been known that one cannot in general first impose the symmetry
and then do the Hamiltonian analysis. That approach happens to
work successfully for Bianchi class A models but does not work for
class B models~\cite{AS91}.

Our energy-momentum is just the value of the Hamiltonian (which
dynamically generates the space-time) with homogeneous boundary
conditions, reference and time slicing. We emphasize that our
approach is entirely geometric (i.e., coordinate independent).
Certainly there are other measures of energy-momentum---and they may
have their own particular virtues---but we do not see how there can
be any other which is more suitable for all the homogeneous
cosmologies.   The results reported here can be considered as a
standard measure of the energy-momentum for homogeneous cosmologies.

\acknowledgments

 We would like to thank C.-M. Chen for discussions and advice and
L. B. Szabados for his helpful remarks.
 TV would like to thank Prof.~H. V. Fagundes for
discussions and the FAPESP, Sao Paulo, Brazil for financial support
during the initial phases of this investigation. JMN and LLS were
supported by the National Science Council of the Republic of China
under the grants NSC 94-2119-M-002-001, 94-2112-M008-038, and
95-2119-M008-027.   JMN was also supported in part by the National
Center of Theoretical Sciences.


\newpage
\begin{thebibliography}{99}

\bibitem{mtw} C.~W. Misner, K.~S. Thorne and J.~A. Wheeler,
{\it Gravitation} (Freeman, San Francisco, 1973) \S 20.4.
\bibitem{tra} A. Trautman, in {\it Gravitation: an Introduction to
      current research}, ed. L.~Witten (Wiley, New York, 1958).
\bibitem{pap} A. Papapetrou, Proc.~R. Ir.~Acad.~A {\bf 25}, 11 (1951).
\bibitem{beto} P.~G. Bergmann and R. Thompson, Phys.~Rev.~{\bf 89}, 400 (1953).
\bibitem{Mol58} C. M{\o}ller, Ann.~Phys.~{\bf 4}, 347 (1958).
\bibitem{Gol58} J.~N. Goldberg, Phys.~Rev.~{\bf 111}, 315 (1958).
\bibitem{lan} L.~D. Landau \&  E.~M. Lifshitz, {\it The
      Classical Theory of Fields} 2nd ed. (Addison-Wesley, Reading, MA, 1973).
\bibitem{wein} S. Weinberg, {\it Gravitation and Cosmology} (Wiley, New York, 1972).
\bibitem{Penrose} R. Penrose, Proc.~Roy.~Soc.~London, A {\bf 381}, 53 (1982).
\bibitem{Sza04} L.~B. Szabados, ``Quasi-local energy-momentum and angular momentum
in GR: A review article'', Living Rev.~Relativity {\bf 7}, 4
 (2004); http://livingreviews.org/lrr-2004-4.
\bibitem{bryo}  J.~D. Brown and J.~W.~York, Jr., Phys.~Rev.~D {\bf 47}, 1407 (1993).
\bibitem{LY03} C.-C.~M. Liu and S.-T. Yau, Phys.\ Rev.\ Lett.~{\bf 90}, 231102 (2003).
\bibitem{berg} G. Bergqvist, Class.~Quantum Grav.~{\bf 9}, 1917 (1992).
\bibitem{PRL99} C.-C. Chang, J.~M. Nester and C.-M. Chen, Phys.\ Rev.\ Lett.~{\bf 83},
1897 (1999).
\bibitem{GC00} C.-M. Chen and J.~M. Nester, Grav.~\& Cosmolo.~{\bf 6}, 257 (2000).
\bibitem{PRD05} C.-M. Chen, J.~M. Nester and R.-S. Tung, Phys.\ Rev.\ D~{\bf 72},
104020 (2005).
\bibitem{Mol61} C. M{\o}ller, Ann.~Phys.~{\bf 12}, 118 (1961).
\bibitem{Nes89} J.~M. Nester, Int.~J. Mod.~Phys.~A {\bf 4},
1755 (1989); Phys.~Lett.~A~{\bf 139}, 112 (1989).
\bibitem{Nes81} J.~M. Nester, Phys.~Lett.~A {\bf 83}, 241 (1981).
\bibitem{Nes84} J.~M. Nester, in {\it Asymptotic
Behavior of Mass and Space-Time Geometry}, Lecture Notes in Physics,
vol 202, ed F.~Flaherty (Springer, Berlin, 1984) pp 155--163.
\bibitem{NT94} J.~M. Nester and R.~S. Tung,
Phys.~Rev.~D~{\bf 49}, 3958 (1994).
\bibitem{Al73} M.~G. Albrow, Nature~{\bf 241}, 56 (1973).
\bibitem{try} E.~P. Tryon, Nature~{\bf 246}, 396 (1973).
\bibitem{Tipler} F. Tipler, Class.~Quantum Grav.~{\bf 2}, L99 (1982).
\bibitem{Mis63} C.~W. Misner, Phys.~Rev.~{\bf 130}, 1590 (1963).
\bibitem{Wal82} R.~P. Wallner, Acta Phys.~Austrica~{\bf 54}, 165 (1982).
\bibitem{coop} F.~I. Cooperstock \& M. Israelit, Gen.~Rel.~Grav.~{\bf 26},
319 (1994).
\bibitem{ros} N. Rosen, Gen.~Rel.~Grav.~{\bf 26}, 319 (1994).
\bibitem{Gar95} J. Garecki, Gen.~Rel.~Grav.~{\bf 27}, 55 (1995).
\bibitem{JKSE}  V.~B. Johri, D. Kalligas, P.~G. Singh \& C.~W.~F. Everitt,
Gen.~Rel.~Grav.~{\bf 27}, 313 (1995).
\bibitem{FD96} S. Feng and Y. Duan, Chin.~Phys.~Lett.~{\bf 13}, 409 (1996).
\bibitem{base} N. Banerjee and S. Sen, Pramana J. Phys.~{\bf 49}, 609 (1999).
\bibitem{rad99} I. Radinschi, Acta Phys.~Slov.~{\bf 49},
789 (1999).
\bibitem{Xu00} S. Xulu, Int.~J. Theor.~Phys.~{\bf 39}, 1153 (2000).
\bibitem{facoo}V. Faraoni and F.~I. Cooperstock, Astrophys.~J.~{\bf 587}, 483 (2003).
\bibitem{var} T. Vargas, Gen.~Rel.~Grav.~{\bf 36}, 1255 (2004).
\bibitem{sovar} L.~L. So and T. Vargas, Chin.~J.~Phys.~{\bf 43}, 901 (2005).
\bibitem{HS79} K. Hayashi and T. Shirafuji, Phys.~Rev.~D~{\bf 19},
3524 (1979).
\bibitem{AGP00} V.~C. de Andrade, L.~C.~T. Guillen and J.~G. Pereira,
     Phys.~Rev.~Lett.~{\bf 84}, 4533 (2000).
\bibitem{KT91} T. Kawai and N. Toma, Prog.~Theor.~Phys.~{\bf 85}, 90 (1991).
\bibitem{Mal} J.~W. Maluf, J. Math.~Phys.~{\bf 35}, 335 (1994); {\bf 36},
4242 (1995); {\bf 37}, 6293 (1996).
\bibitem{BV01} M.~Blagojevi\'c and M.~Vasili\'c, Phys.~Rev.~D~{\bf 64}, 044010 (2001).
\bibitem{OR06} Y.~N. Obukhov and  G.~F. Rubilar, Phys.~Rev.~D~{\bf 73},
124017 (2006).
\bibitem{ellis} G.~F.~R. Ellis and M.~A.~H. MacCallum,
     Comm.\ Math.\ Phys.~{\bf 12}, 108 (1969).
\bibitem{kra} D.~Kramer, H.~Stephani, E.~Herlt and E.~Schmutzer, {\it Exact
      solutions of Einstein's field equations} (Cambridge University Press, Cambridge, 1980).
\bibitem{Sza92} L.~B. Szabados, Class.~Quantum
Gavit.~{\bf 9}, 2521 (1992).
\bibitem{Nes04} J.~M. Nester, Class.~Quantum Gravit.~{\bf 21}, S261 (2004).
\bibitem{gensource} The result
actually holds for quite general sources, e.g., spinors etc., but
the derivation for such cases is more complicated, and not so
enlightening.
\bibitem{JR06} M. Janssen and J. Renn,
``Untying the Knot: How Einstein Found His Way Back to Field
Equations Discarded in the Zurich Notebook'', in {\it The Genesis of
General Relativity} Vol. 2, {Einstein's Zurich Notebook: Commentary
and Essays} ed J.~Renn (Springer, 2006) pp 849--925.

\bibitem{Hehl} F.~W. Hehl, ``Four Lectures on Poincar\'e gauge field theory'',
in {\it Proceedings of the 6th Course on Spin, Torsion, Rotation,
and Supergravity} Erice, eds P.~G.~Bergmann, V.~de~Sabbata (Plenum
New York, 1980) p 5.
\bibitem{CQG99} C.-M. Chen and J.~M. Nester, Class.~Quantum Grav.~{\bf 16}, 1279 (1999).
\bibitem{Mal+} J.~W. Maluf and J.~F. da Roche-Neto, J. Math.~Phys.~{\bf 40},
1590 (1999).
\bibitem{Mal+02} J.~W. Maluf, J.~F. da Roche-Neto, T.~L.`M. Toribio
and K.~M. Castllo-Blanco, Phys.~Rev.~D~{\bf 65}, 124001 (2002).
\bibitem{exceptions}Actually certain teleparallel
analogues of the Einstein, Landau-Lifshitz and Bergman-Thompson
pseudotensors were recently defined in~\cite{var}; however they have
not so far been given serious consideration and study.
\bibitem{KBL97} J. Katz, J. Bi{\v c}{\'a}k \&
D. Lynden-Bell, Phys.~Rev.~D~{\bf 55}, 5957 (1997).
\bibitem{linearmom}$P_c$ is the conserved quantity associated with
displacements having constant components along the spatial legs of
the preferred tetrad. As these spatial displacements do not commute
it is not so clear to what degree these conserved  quantities
actually resemble linear momentum.
\bibitem{AS91} A. Ashetkar and J. Samuel, Class.~Quantum Grav.~{\bf 8}, 2191 (1991).
\bibitem{MPLA7} C.-M. Chen, J.-L. Liu and J.~M. Nester,
Mod.\ Phys.\ Lett.~A~{\bf 22}, 2039 (2007).
\bibitem{NCL07}  J.~M. Nester, C.-M. Chen and J.-L. Liu, ``Quasi-local energy for
cosmological models'', in {Proceedings of the Eleventh Marcel
Grossmann Meeting on General Relativity} (World Scientific,
Singapore, 2007), in press.

\bibitem{ADM60} R. Arnowitt, ~S. Deser, and
C.~W. Misner, Ann.~Phys.~{\bf 11}, 116 (1960).
\bibitem{BD68} D.~R. Brill and S. Deser, Ann.~Phys.~{\bf 50}, 548 (1968).
\bibitem{Ger78} R. Geroch, ``The positive-mass conjecture'',
in {\it Theoretical Principles in Astrophysics and Relativity} eds.
N.~R. Lebovitz, W.~H. Reid and P.~O. Vandervoort (University of
Chicago Press, 1978) p 245.
\bibitem{BJ80} D.~R. Brill and P.~S. Jang, in {\it General Relativity and Gravitation,} Vol. 1., ed A.~Held (Plenum, New
York, 1980) p 173.
\bibitem{Hor84} G.~T. Horowitz, in {\it Asymptotic
Behavior of Mass and Space-Time Geometry}, Lecture Notes in Physics,
vol 202, ed F.~Flaherty (Springer, Berlin, 1984) pp 1--20.
\bibitem{LY06} C.-C.~M. Liu and S.-T Yau, J. Amer.\ Math.\ Soc. {\bf 19}, 181 (2006).

\end{thebibliography}
\end{document}